\documentclass[10pt]{iopart} 
\usepackage{graphicx}
\usepackage{epstopdf}
\usepackage{siunitx}
\usepackage{cite}
\usepackage{hyperref}
\usepackage{color}

\renewcommand{\textcolor}[1]{}

\begin{document}

\title{Superconducting Pb stripline resonators in parallel magnetic field and their application for microwave spectroscopy}

\author{Nikolaj G Ebensperger, Markus Thiemann, Martin Dressel and Marc Scheffler}

\address{1. Physikalisches Institut, Universit{\"a}t Stuttgart, Pfaffenwaldring 57, D-70569 Stuttgart, Germany}
\ead{scheffl@pi1.phyisk.uni-stuttgart.de}
\vspace{10pt}
\begin{indented}
\item[]October 10th, 2016
\end{indented}

\begin{abstract}
Planar superconducting microwave resonators are key elements in a variety of technical applications and also act as sensitive probes for microwave spectroscopy of various materials of interest in present solid state research. Here superconducting Pb is a suitable material as a basis for microwave stripline resonators.

To utilize Pb stripline resonators in a variable magnetic field (e.g. in ESR measurements), the electrodynamics of such resonators in finite magnetic field has to be well understood. Therefore we performed microwave transmission measurements \textcolor{blue}{(with ample applied power to work in linear response)} on superconducting Pb stripline resonators in a variable, parallel magnetic field. We determined surface resistance, penetration depth as well as real and imaginary parts, $\sigma_1$ and $\sigma_2$, of the complex conductivity of superconducting Pb as a function of magnetic field. Here we find features reminiscent of those in temperature-dependent measurements, such as a maximum in $\sigma_1$ (‘coherence peak’). At magnetic fields above the critical field of this type-I superconductor we still find \textcolor{blue}{a} low-loss microwave response, which we assign to remaining superconductivity in the form of filaments within the Pb. Hysteresis effects are found in the quality factor of resonances once the swept magnetic field
has exceeded the critical magnetic field. This is due to normal conducting areas that are pinned and can therefore persist in the superconducting phase. Besides zero-field-cooling we show an alternative way to eliminate these even at $T<T_c$. Based on our microwave data, we also determine the critical magnetic field and the critical temperature of Pb in a temperature range between 1.6K and 6.5K and magnetic fields up to 140mT, showing good agreement with BCS predictions. We furthermore study a Sn sample in a Pb resonator to demonstrate the applicability of superconducting Pb stripline resonators in the experimental study of other (super-)conducting materials in a variable magnetic field.
\end{abstract}

\ioptwocol

\section{Introduction}

Building a microwave device (or just a simple element of certain geometry) from a superconducting material and obtaining detailed knowledge of its properties is useful in several different contexts. Firstly, it can help for the full understanding of the superconducting material itself. Microwave frequencies, with a typical range of \SI{1}{}-\SI{100}{GHz}, correspond to thermal energies of \SI{0.05}{}-\SI{5}{K} and are in the range of or below critical temperatures $T_c$ and energy gaps of many superconductors. Thus the microwave response of a superconductor depends on both the dissipationless Cooper pairs and the dissipating thermally excited quasiparticles \cite{dressel_buch,Pracht2013}. Accordingly, microwave experiments probe the different charge carriers, and give access to quantities such as superconducting penetration depth (related to Cooper pair density and superfluid stiffness), energy gap, surface resistance, and complex optical conductivity \cite{dressel_microwave_sigma,kobayashi_microwave_q_sigma}, but they can also reveal different dynamical properties, such as quasiparticle relaxation, critical fluctuations, or collective excitations \cite{Feller2002,Turner2003,Kitano2006,Steinberg2008,Cea2014,Sherman2015}.

Another field of interest for superconducting microwave devices are technical applications such as microwave filters, but they also include numerous types of resonators, detectors, amplifiers, and related devices in fundamental research \cite{Day2003,Wallraff2004,Goeppl2008,Natarajan2012}. \textcolor{blue}{Depending on application, e.g.\ in astrophysics or quantum information science, planar superconducting resonators might be operated with vastly different power levels, and, correspondingly, different microscopic processes govern the resonator performance and possible strategies for device improvements \cite{OConnell2008,Barends2010,Macha2010,Megrant2012,deVisser2013}.}
Microwave spectroscopy in condensed matter physics is yet another resonator application. For many materials of present research, such as magnetic and heavy-fermion materials, there exist important low energy scales that can be accessed with microwave spectroscopy \cite{Schwartz2000,Schwarze2015,scheffler_heavy_fermions,Awasthi1993,Scheffler2005b,scheffler_heavy_fermions_2}, and a very prominent case here is the vast range of conventional and unconventional superconductors, including cuprate, pnictide, organic, heavy-fermion, and strongly disordered superconductors \cite{Hashimoto2009,Truncik2013,Liu2013,Beutel2016}. To this end, a variety of different microwave spectroscopy techniques have been developed \cite{Scheffler2015,Booth1994,Turner2004,Scheffler2005a,Huttema2006,Pompeo2007}. 

Studying conducting materials is also our motivation for the development of superconducting Pb stripline resonators that can act as compact and sensitive probes in cryogenic microwave spectroscopy \cite{DiIorio1988,scheffler_stripline,hafner}. Here Pb with $T_c\approx\SI{7,2}{K}$ and a critical magnetic field of about $B_c\approx\SI{80}{mT}$ is a well suited basis material \cite{kittel}, as it can be processed for thin films rather easily. 
For the spectroscopic study of certain materials \cite{scheffler_heavy_fermions,Schwarze2015,Sichelschmidt2010,clauss_esr,wiemann_esr}, application of a magnetic field in the order of tens of mT is helpful, and therefore we address the microwave properties of Pb stripline resonators in finite magnetic field. This is not only relevant for spectroscopy, but like any exposure of a superconductor to a magnetic field is also an interesting issue on its own. Operating planar microwave devices in magnetic field has been studied in the context of vortex motion, dissipation, and pinning in type-II superconductors, where a pronounced field dependence is found that is unwanted for applications \cite{Song2009,Bothner2012,bothner_demagnitization,steele_molybden}. In contrast, for type-I superconductors such questions have, so far, attracted much less attention. In fact, one motivation for our work on Pb resonators is that for a type-I superconductor, up to the critical field, there should be very weak field dependence, which is convenient for spectroscopic studies.
However, in earlier work it has been found that \textcolor{blue}{normal-conducting} areas similar to vortices can persist in bulk superconductors of type I \cite{song_vortices_in_re_al} and especially in Pb \cite{ge_flux_patterin_in_lead,alers_intermediate_state_pb,velez_magnetic_lead,velez_temp_magn_lead}. This flux trapping effect, as well as \textcolor{blue}{an} increased critical magnetic field for resistance restoration and a rather broad resistance transition commonly occur in ``hard'' superconductors of type I that contain many impurities and imperfections \cite{schoenberg_superconductivity_book,seraphim_resistance_in_Ta,budnick_resistance_in_tantalum}.
However, this has so far rarely been investigated with microwaves in a variable magnetic field, yet a detailed understanding is required as residual \textcolor{blue}{normal-conducting} areas in Pb could gravely influence the electrodynamic properties.

In this work we examine Pb stripline resonators in varying external magnetic field up to \SI{140}{mT}, below and above the critical magnetic field $B_c$, at temperatures of \SI{1,6}{K} to \SI{6,5}{K}.
With the determination of penetration depth and surface resistance we are able to access the complex conductivity and find striking similarities to measurements in variable temperature.
We study possible influences of remaining \textcolor{blue}{normal-conducting} areas in the superconducting phase by first suppressing superconductivity with magnetic fields above $B_c$ and then regaining superconductivity by decreasing the magnetic field.
In order to demonstrate the possibility to use Pb stripline resonators for microwave spectroscopy applications in varying magnetic field, we study a Sn sample with a critical temperature of $T_c\approx\SI{3,7}{K}$ and a critical magnetic field of $B_c\approx\SI{30}{mT}$ \cite{kittel} with the stripline resonator.

\section{Experimental setup and methods}
\begin{figure}[h]
\centering
\includegraphics[width=\linewidth]{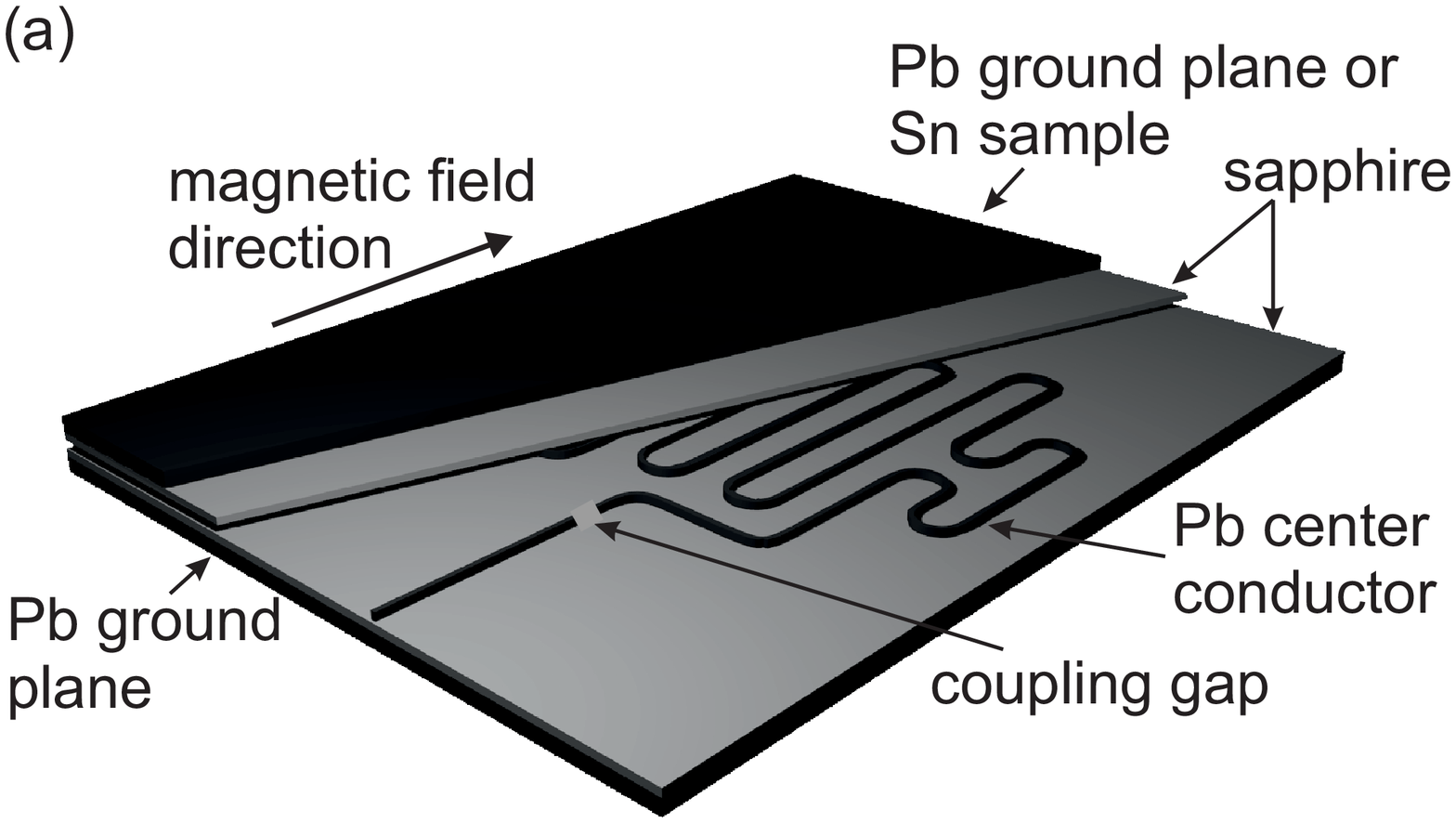}
\includegraphics[width=\linewidth]{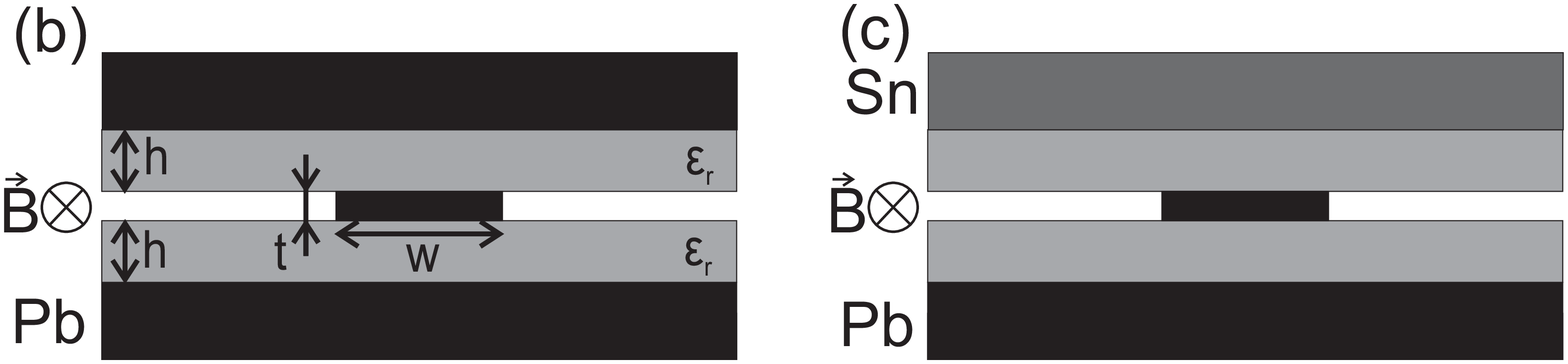}
\caption{Schematic geometry of a microwave stripline resonator in 3D view (a) and in cross section without sample (b) and with additional Sn sample (c). As indicated, the direction of the magnetic field is parallel to the resonator.}
\label{fig:geometry}
\end{figure}

The geometry of our microwave stripline resonators is shown in Fig.~\ref{fig:geometry}, as three-dimensional schematic view (a) and in cross section (b,c). They consist of a \SI{1}{\micro\meter} thick and \SI{45}{\micro\meter} wide center conductor, which is embedded between two dielectric sapphire plates of \SI{127}{\micro\meter} thickness, which in turn are sandwiched between two conductive ground planes above and below. The Pb center conductor is evaporated onto one of the sapphire plates and has a meander structure.

Two gaps in the center conductor separated from each other at a certain distance $l$ reflect the microwave signal to a large extend, allowing the microwave to form a standing wave between the gaps. Fundamental frequencies vary with the separation $l$ of the gaps and in our case range between \SI{1}{GHz} to \SI{2}{GHz} with higher harmonics of the fundamental frequency being measurable up to \SI{20}{GHz}.
This propagation of the microwaves allows measurements with superconducting stripline resonators to be particularly sensitive to the state of the material near the surface.
In order to study an additional superconducting material (here: Sn), one of the ground planes is replaced by a sample, as depicted in Fig.~\ref{fig:geometry}(c).

\textcolor{blue}{Depending on the field of application in which superconducting microwave resonators are used, the power that is applied to drive them has to be adjusted properly. Circuit QED experiments in quantum information science, for example, might employ an average number of microwave photons in the resonator that can be much smaller than one \cite{Wallraff2004} whereas the drive/readout powers for superconducting resonators can be many orders of magnitude higher for other applications \cite{Schuster2005}. At very small power, the resonator response might be non-linear, i.e.\ depend on the applied microwave power, due to two-level fluctuators \cite{OConnell2008} whereas at very high powers there can be non-linear behavior due to the superconductor \cite{Chin1992}.
For our experiments, we want the applied power to be high enough for good signal-to-noise ratio in the transmission measurement but yet well below the power range where our stripline resonators becomes non-linear,\cite{hafner}, and we eventually used an input power of \SI{-30}{dBm}.}

At resonance frequencies of the stripline resonator the microwave meets the resonance condition resulting in a maximum of microwave transmission. At frequencies deviating from the resonance frequency the transmission decreases rapidly, leading to a Lorentzian form of the transmission signal around resonance frequencies \cite{petersan_resonator_general}. After correction of the data performed similar to \cite{hafner}, a fit of a Lorentzian with center frequency $\nu$ and full width at half maximum $\Delta\nu$ is possible. This enables the calculation of the quality factor $Q = \nu/\Delta\nu$ for any given resonance, defined as the ratio of power stored to power dissipated each cycle. It is a measure of losses in the resonator and dominantly influenced by the conductors. By replacing one ground plane with a superconducting sample of different material that has lower critical temperature and lower critical magnetic field than the underlying resonator, the losses in the resonator become dominated solely by the losses in the sample, giving experimental access for measurements explicitly on the sample.

The determination of the quality factor $Q$ also allows for the calculation of the surface resistance \cite{hafner} 
\begin{equation}
R_s = \pi \mu_0 \Gamma \nu/Q \, , \label{eq:rs}
\end{equation}
with $\Gamma$ a geometrical constant of the resonator \cite{hafner,wheeler} and $\mu_0$ the permeability constant. Additionally the surface reactance \cite{dressel_buch}
\begin{equation}
X_s = -Z_0\nu \lambda /(2\pi c) \label{eq:xs}
\end{equation}
can be calculated with $\lambda$ the penetration depth into the superconducting material and $Z_0\approx\SI{377}{\ohm}$ the impedance of the vacuum. Calculation of the real and imaginary parts of the complex conductivity $\hat{\sigma} = \sigma_1 + i\sigma_2$ is then possible using
\begin{equation}
\sigma_1 = -Z_0 2\omega (X_s R_s)/(c(X_s^2+R_s^2)^2) \label{eq:sigma1}
\end{equation}
and
\begin{equation}
\sigma_2 = Z_0 \omega(X_s^2-R_s^2)/(c(X_s^2+R_s^2)^2) \, , \label{eq:sigma2}
\end{equation}
with $c$ the speed of light.

We studied the stripline resonator experimentally with a vector network analyzer~(VNA) in transmission, recording the complex transmission parameter S$_{21}$. We performed the measurements in a $^4$He cryostat, reaching down to temperatures of about \SI{1,6}{K}. A static magnetic field was applied parallel to the resonator as indicated in Fig. \ref{fig:geometry} \footnotemark. Before each measurement with variable magnetic field, zero field cooling~(ZFC) has been performed, meaning the resonator has been cooled down below its superconducting transition temperature in the absence of a magnetic field and the magnetic field has been applied subsequently at a fixed temperature.

\section{Results and discussion}
\begin{figure}[h]
\centering
\includegraphics[width=\linewidth]{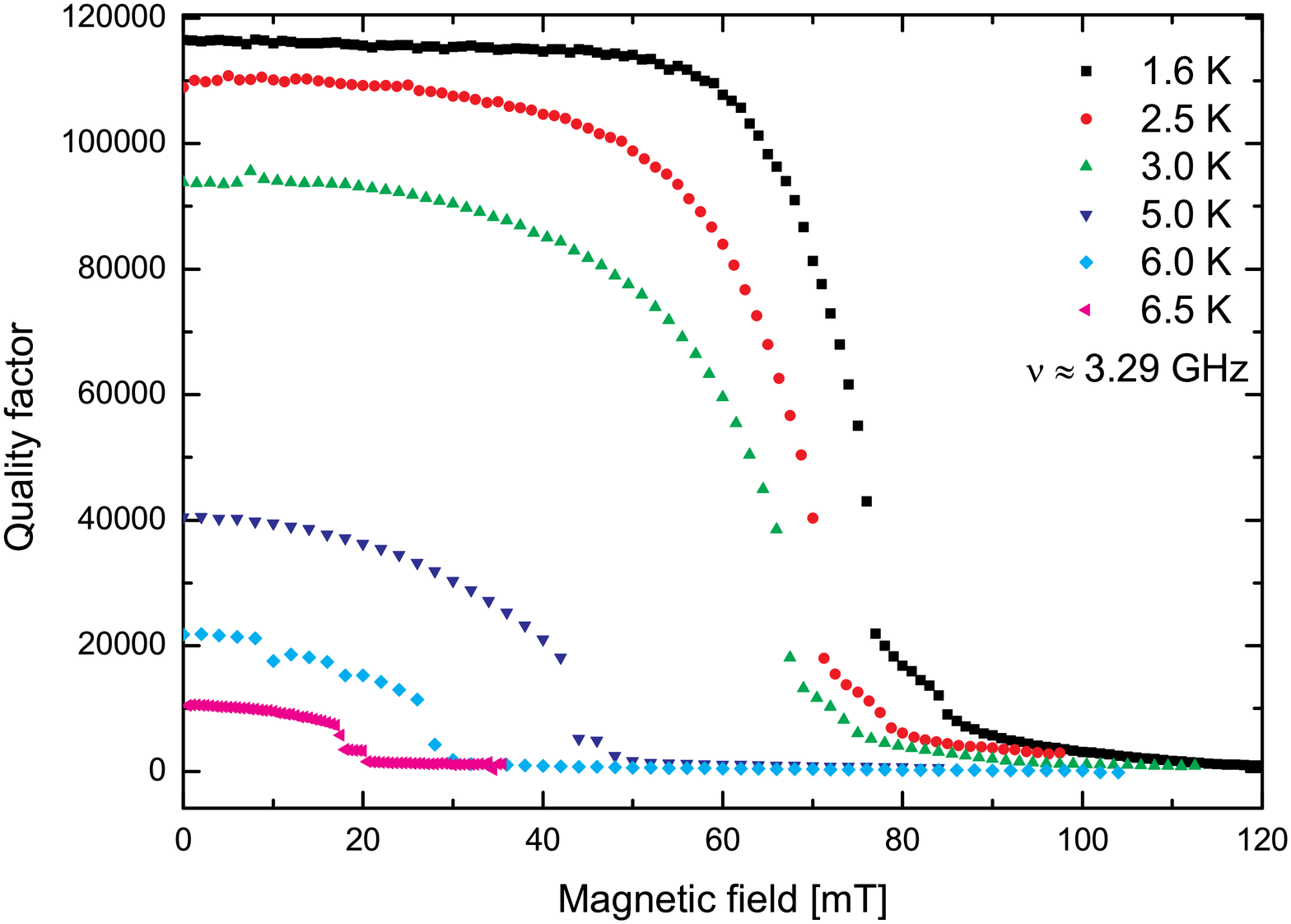}
\caption{Quality factor $Q$ for a resonance of the Pb resonator with increasing magnetic field up to \SI{120}{mT} at different temperatures in the range between \SI{1,6}{K} and \SI{6,5}{K}. $Q$ shows only a small decrease in the Meissner state with $B<B_c(T)$. Close to $B_c(T)$, $Q$ decreases drastically before entering the intermediate state.  $Q$ still shows a smaller but substantial remaining value even at $B>B_c(T)$, indicating possible superconductivity in the form of filaments.}
\label{fig:bleiq}
\end{figure}

For a resonator made entirely out of Pb (center conductor as well as ground planes) the quality factor $Q$ in a varying magnetic field has been determined as shown in Fig \ref{fig:bleiq} for several different temperatures. $Q$ was obtained for the second mode of the resonator at \SI{3,29}{GHz} with a magnetic field up to \SI{120}{mT}. At fields below the critical field $B_c(T)$, $Q$ only decreases little.
This regime resembles the Meissner state.
Following a drastic decrease, indicating increasing losses close to the normal conductor transition, a plateau emerges leading to two kinks in the course of $Q$ (e.g. in the range of \SI{77}{}-\SI{84}{mT} at $T=\SI{1,6}{K}$). This area is the intermediate state where superconducting and normal conducting areas coexist \cite{alers_intermediate_state_pb}. Since a perfectly parallel alignment of the resonator to the magnetic field is practically impossible, an intermediate phase is always existent \cite{reeber_magnetic_field_orientation} and was found throughout all our measurements. This has also been found in earlier publications on Pb microwave resonators \cite{scheffler_heavy_fermions,koepke_neustes}, but was not discussed there.

Leaving this state \textcolor{blue}{towards} higher fields, $Q$ becomes comparably small, however it does not vanish completely. This could very well be caused by remaining superconductivity in the resonator, which is typical for hard superconductors of type I \cite{schoenberg_superconductivity_book}. Several similar effects are found, e.g. in \textcolor{blue}{tantalum} where a gradual transition of the resistance at $B>B_c$ happens and is associated with superconducting ``filaments'' \cite{seraphim_resistance_in_Ta,budnick_resistance_in_tantalum}.
This gradual transition has also been theoretically predicted \cite{saint-james_hc3_theoretical_prediction} and found in 
alloys of type II \cite{cardona_surface_supercond,hempstead_hc3_in_typeII}.
In our measurements these effects lead to the small values of $Q$, often persisting to magnetic fields way above $B_c(T)$, e.g. in the measurement  at \SI{1,6}{K} up to about \SI{120}{mT}, while $B_c(T=\SI{1,6}{K})\approx\SI{77}{mT}$, showing that our Pb specimen contains impurities and imperfections making it a hard superconductor.

\begin{figure}[h]
\centering
\includegraphics[width=\linewidth]{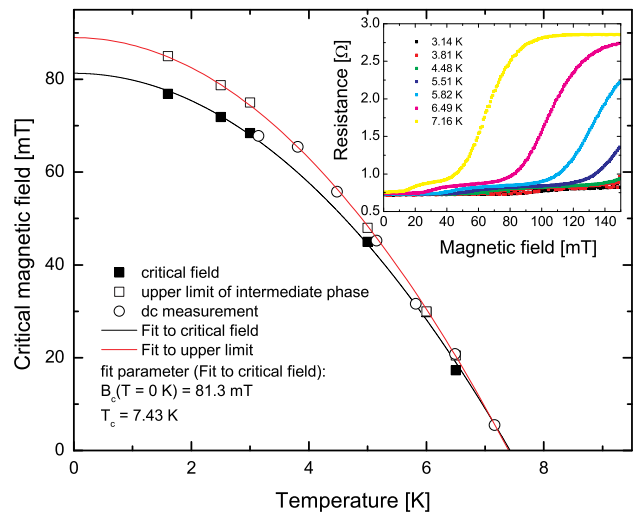}
\caption{$B_c(T)$ and upper limit of the intermediate phase of Pb in dependence of temperature as determined from microwave measurements (squares). A \textcolor{blue}{DC measurement} of the resistivity in magnetic field (using a transmission line with the same geometry as the resonator without the gaps) is shown in the inset. $B_c(T)$ derived at the first increase in resistivity is shown in the main panel as empty circles. The empirical equation for $B_c(T)$ has been fitted to the microwave data and the values for $T_c$ and $B_c(0)$ show good agreement with earlier measurements \cite{crescenzo_pb_magn_temp_thickness,cody_pb_magn_thickness,indovina_pb_crit_magn_microwave}.}
\label{fig:bleicritmagn}
\end{figure}

From the data in Fig.~\ref{fig:bleiq} the critical magnetic field of Pb could be determined for each temperature at half the maximum value of $Q$ as well as the upper field limit of the intermediate state. This is shown in Fig.~\ref{fig:bleicritmagn}. Additionally DC-measurements using the same geometry as the resonator (without the gaps for microwave reflectance) are shown in the inset. Here $B_c(T)$ has been determined at the first increase in resistivity. The subsequent course of the resistivity at higher fields shows a gradual increase typical for hard superconductors and the existence of superconducting filaments at $B>B_c(T)$ \cite{schoenberg_superconductivity_book,seraphim_resistance_in_Ta,budnick_resistance_in_tantalum}.
The $B_c(T)$ data of the microwave measurement has been fitted with the empirical equation 
\begin{equation}
B_c(T) = B_c(0)(1-(T/T_c)^2) \, , \label{eq:bcsfit}
\end{equation}
with $B_c(T)$ the critical field at a given temperature and $B_c(0)$ the critical field at zero temperature.
The resulting values $B_c(0) = \SI{81,3}{mT}$ and $T_c = \SI{7,43}{K}$ are in good agreement with earlier measurements \cite{crescenzo_pb_magn_temp_thickness,cody_pb_magn_thickness,indovina_pb_crit_magn_microwave}.

\begin{figure}[h]
\centering
\includegraphics[width=\linewidth]{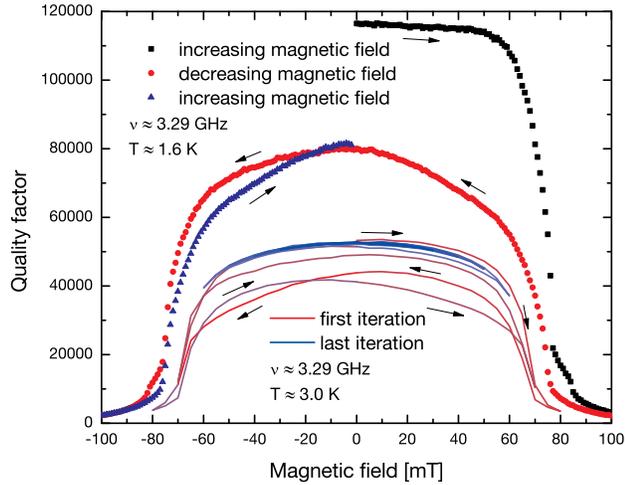}
\caption{Filled symbols show the quality factor for a resonance of the Pb resonator with variable magnetic field up to \SI{100}{mT} after performing ZFC. The magnetic field has first been increased to a positive maximum value and afterwards decreased to a negative maximum value. $Q$ shows hysteresis both for positive magnetic fields and for negative magnetic fields as it does not reach initial values after once exceeding $B_c(T)$. Lines show $Q$ of the Pb resonator for variable magnetic field alternating between positive and negative maximum values with decreasing maximum value of the swept field for consecutive iterations. After showing strong hysteresis in the first iterations, $Q$ rises almost to initial values with following iterations, indicating annihilation of normal conducting areas.}
\label{fig:hysteresis}
\end{figure}

Strong hysteresis could be seen when increasing the external magnetic field to values higher than $B_c(T)$. $Q$ for a resonance at \SI{3,29}{GHz} with an increasing and afterwards decreasing field is shown as filled symbols in Fig~\ref{fig:hysteresis} at $T\approx\SI{1.6}{K}$. Before this measurement zero field cooling (ZFC) was performed, to allow for full restoration of superconductivity and maximum $Q$.
During the increase of the field $Q$ drops to low values similar to Fig~\ref{fig:bleiq} at fields higher than $B_c(T)$. When decreasing the magnetic field again\textcolor{blue}{, $Q$} starts to rise again at similar $B_c(T)$, however it does not reach its initial values of about \SI{1,16e5}{}, but instead only reaches a maximum value of about \SI{7,97e4}{} at $B=0$. Additionally $Q$ still increases substantially below $B_c(T)$ with decreasing field compared to the only small change during the initial field increase. Increasing the field to negative values and exceeding $B_c(T)$ shows similar behavior concerning hysteresis although lower in magnitude.

Similar to the superconducting filaments at $B>B_c(T)$, these hysteresis effects are caused by normal conducting filaments trapped in surrounding superconducting material at $B<B_c(T)$. 
Impurities within the Pb of the resonator pin the normal conducting areas, originating from the intermediate state.
This is often characteristic for hard superconductors \cite{schoenberg_superconductivity_book}. In Pb similar effects have already been shown earlier \cite{desorbo_intermediate_state,ge_flux_patterin_in_lead,alers_intermediate_state_pb,velez_magnetic_lead,velez_temp_magn_lead} and we now confirm these using microwave measurements with stripline resonators.
These effects have to be considered when performing microwave studies on Pb resonators as they can gravely affect measurements. Zero-field-cooling before each measurement is therefore a well-fitting procedure to circumvent these unwanted effects.

The lines in Fig~\ref{fig:hysteresis} show a measurement using the same resonator and mode at \SI{3,29}{GHz} now at $T\approx\SI{3.0}{K}$. The external magnetic field has been repeatedly increased to a maximum positive value and afterwards decreased to a maximum negative value. The first maximum field exceeded $B_c(T)$ and thus $Q$ again shows hysteresis. The following iterations of this measurement however, lowered the maximum field each time to a value lower than the previous iteration (both positive and negative). Since $B_c(T)$ is not exceeded during later iterations, $Q$ increases again, reaching almost initial values. This is explained by the introduction of normal conducting areas of opposite polarity into the resonator with the second half of each iteration, where the magnetic field has negative value \cite{bothner_demagnitization}. Normal conducting areas of opposing polarity annihilate leading to a reduction of overall normal conducting area.

The regaining of the original state with this procedure can be useful if ZFC can not be performed (e.g. if it is unpractical to exceed $T_c$), since a simple magnetic field sweep procedure can allow for complete restoration even at very low temperatures.

\begin{figure*}[tbph]
\centering
\includegraphics[width=0.9\linewidth]{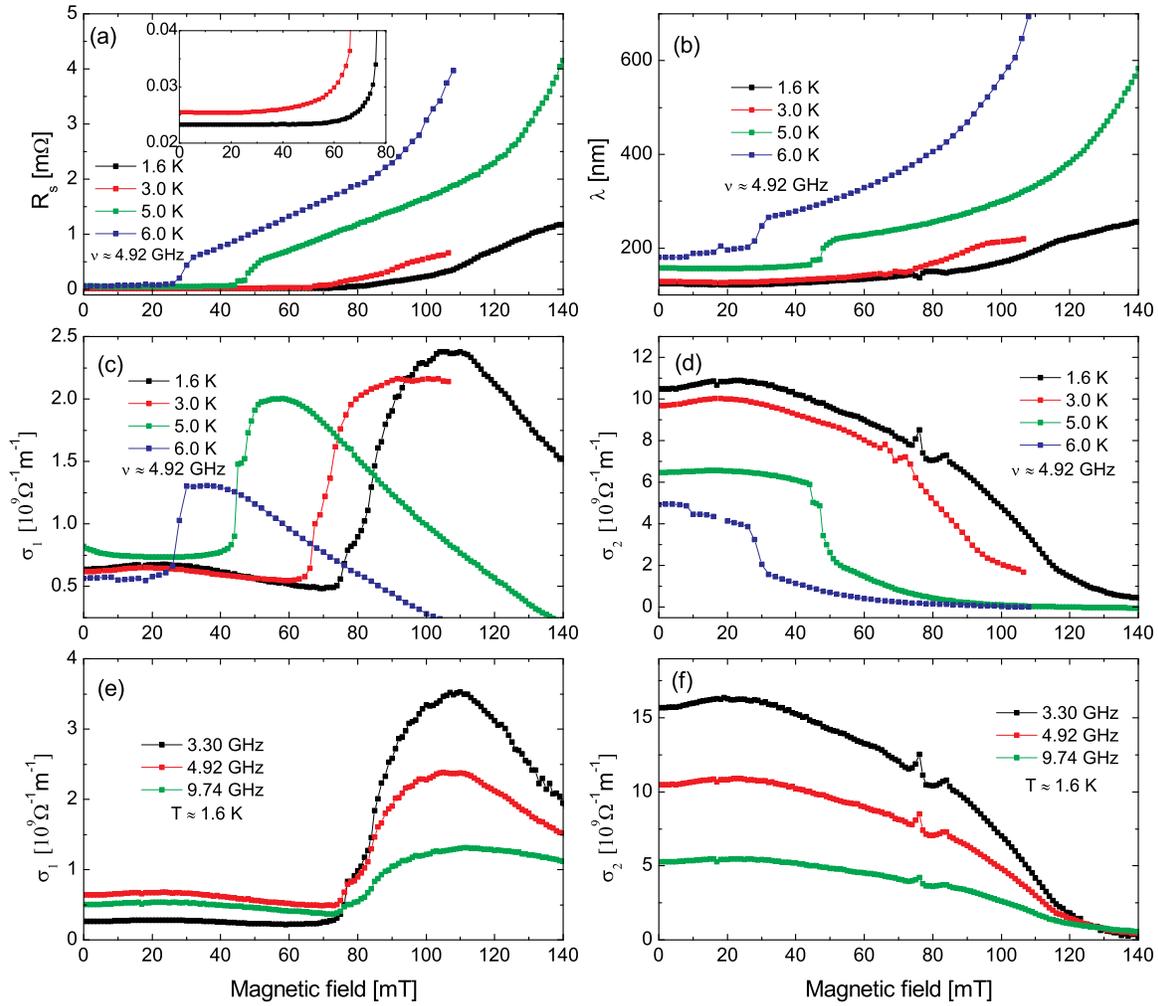}
\caption{(a) Surface resistance $R_s$ and (b) penetration depth $\lambda$ for a Pb resonator at different temperatures plotted against the magnetic field. At $B>B_c(T)$ only superconducting filaments remain in the resonator increasing $R_s$ and $\lambda$ heavily. Both quantities also increase with temperature. (c) real part of the complex conductivity $\sigma_1$ and (d) imaginary part $\sigma_2$. Both show similar shape to temperature-dependent measurements, but do not vanish immediately at $B>B_c(T)$ indicating superconducting filaments as well. (e) and (f) show $\sigma_1(B)$ and $\sigma_2(B)$, respectively, for different resonator frequencies.}
\label{fig:sigma}
\end{figure*}

With increasing temperature the resonance frequency $\nu$ of each resonance shifts to lower frequencies. This is due to the penetration depth increasing with temperature \cite{bcs_theory}, which leads to a change of the effective geometry of the resonator. The frequency decreases following the expression \cite{hafner}
\begin{equation}
\nu = \nu_0 \left[1+\lambda(T) \pi \mu_0 /\Gamma\right]^{-1/2} \, , \label{eq:resfreq}
\end{equation}
with $\nu_0$ the resonance frequency at zero penetration depth. $\lambda(T)$ is the temperature-dependent penetration depth, \textcolor{blue}{which we can assume to change as} $\lambda = \lambda_0 \left(1-(T/T_c)^4\right)^{-1/2}$ near $T_c$ \cite{tinkham_introduction}. By fitting (\ref{eq:resfreq}) to a shift of a resonance with rising temperature, $\nu_0$ of the unpenetrated stripline is acquired. $\nu_0$ is now used to calculate $\lambda(B)$ by assuming the penetration depth magnetic field dependent in equation~(\ref{eq:resfreq}) and solving for $\lambda(B)$ leading to $\lambda(B) = \Gamma\left[\left(\nu_0/\nu(B)\right)^2-1\right]$. The resulting $\lambda(B)$ is shown in Fig.~\ref{fig:sigma}(b). It increases with rising magnetic field. This increase is comparably small for magnetic fields $B<B_c(T)$, and increases heavily at magnetic fields $B>B_c(T)$, where only superconducting filaments remain. $\lambda$ also shows temperature dependence, increasing with higher temperature, as expected \cite{hafner}.

\textcolor{blue}{Subsequent} calculation of surface resistance $R_s$ and surface reactance $X_s$ in dependence of magnetic field is possible. The exact calculation of $R_s$ and $X_s$ was performed similar to \cite{hafner} and \cite{dressel_buch} using equations (\ref{eq:rs}) and (\ref{eq:xs}). This is possible since the energies of the microwaves we used are far lower than the energy gap of superconducting lead \cite{glaever_energy_gap,richards_energy_gap}. $R_s$ is displayed in Fig.~\ref{fig:sigma}(a). It shows only marginal increase up to $B_c(T)$ (see inset in Fig.~\ref{fig:sigma}(a)), the shape of which resembles measurements in variable temperature very well, with the magnitude also being similar \cite{hafner}. At $B_c(T)$, $R_s$ leaps to higher values and continues rising with increasing field, indicating increasing losses in the resonator when only superconducting filaments remain.

Real and imaginary part of the complex conductivity can be \textcolor{blue}{determined} using $R_s$ and $X_s$ \cite{dressel_buch} with equations (\ref{eq:sigma1}) and (\ref{eq:sigma2}). The real part $\sigma_1$ of the lead resonator for different temperatures ranging between \SI{1,6}{K} and \SI{6,0}{K} is depicted in Fig.~\ref{fig:sigma}(c), the imaginary part $\sigma_2$ in Fig.~\ref{fig:sigma}(d). Looking at several higher harmonics of the Pb resonator, frequency-dependent $\hat{\sigma}$ was calculated and its real and imaginary parts for different frequencies are depicted in Fig~\ref{fig:sigma}(e) and Fig.~\ref{fig:sigma}(f).
$\sigma_1$ shows finite value at magnetic fields lower than $B_c(T)$. Close to and above $B_c(T)$ it increases rapidly and forms a peak similar to a coherence peak seen in temperature-dependent $\sigma_1$ measurements \cite{hafner,dressel_edyn_supercond,holczer_coherence_peak}. The coherence peak in temperature-dependent measurements however, is always located below $T_c$, whereas the peaks we measured are located above $B_c$.
At even higher fields, $\sigma_1$ decreases again. Similar to the temperature dependent $B_c(T)$ from Fig.~\ref{fig:bleicritmagn}, the peak shifts to lower fields with increasing temperature and also decreases in magnitude.

$\sigma_2$ (Fig.~\ref{fig:sigma}(d,f)) shows high values at low magnetic fields and a continuous decrease towards higher fields. At $B>B_c(T)$ the change in value gets more pronounced, approaching $\sigma_2(B)\approx\SI{0}{}$ at fields well above $B_c(T)$. The general shape of the course again resembles the shape of measurements in variable temperature well, although $\sigma_2(T)\approx\SI{0}{}$ already at $T\approx T_c$.

The non-vanishing values of $\sigma_1$ and $\sigma_2$ at fields higher than $B_c(T)$ are caused by the remaining superconducting filaments as well. Additional kinks in the course of $\sigma_1$ and $\sigma_2$ are located at magnetic fields coinciding with the intermediate phase (e.g. at \SI{77}{mT} to \SI{84}{mT} at \SI{1,6}{K}).

Both $\sigma_1$ as well as $\sigma_2$ show frequency-dependent behavior (see Fig.~\ref{fig:sigma}(e,f)), as they decrease with an increase in frequency. This is in good agreement with earlier work as both surface resistance and surface reactance increase with frequency \cite{hafner,dressel_edyn_supercond,thiemann_sigma}.

\begin{figure}[h]
\centering
\includegraphics[width=\linewidth]{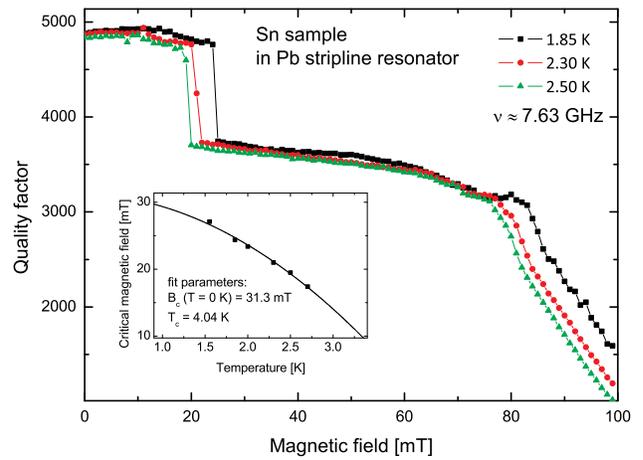}
\caption{Quality factor $Q$ for a Pb stripline resonator with a Sn sample in dependence of external magnetic field. $Q$ shows an initial sharp decrease at magnetic fields of about \SI{19}{mT} to \SI{24}{mT}, which is due to the sample transitioning into the normal conducting state, enhancing losses in the resonator. Inset:~$B_c(T)$ of Sn determined from the initial decrease of $Q$ fitted with equation~(\ref{eq:bcsfit}). The values for $B_c(0)$ and $T_c$ are in good agreement with literature \cite{onori_sn_crit_magn,kittel}.}
\label{fig:zinnprobe}
\end{figure}

By replacing one ground plane of the Pb stripline resonator with a Sn sample (made of Sn foil) we are able to access the superconducting \textcolor{blue}{properties of Sn}. As the electromagnetic losses in the resonator become greatly influenced by the sample, the overall value of $Q$ is strongly affected, showing drastically lower values than an all-Pb resonator (maximum of about \SI{5e3}{}, opposed to all-Pb with about \SI{1,2e5}{}).
$Q$ also shows an additional decrease at $B_c(T)$ of Sn as seen in Fig.~\ref{fig:zinnprobe} (here at about \SI{19}{mT} to \SI{24}{mT}). This is due to \textcolor{blue}{the transition of the sample} from the superconducting to the normal conducting state, leading to increased losses in the resonator. A shift of this drop to lower magnetic fields with increasing temperature is seen. This can be used to extract $B_c(T)$ and fit equation~(\ref{eq:bcsfit}) to the data. The values are depicted in the inset of Fig.~\ref{fig:zinnprobe} and result in $B_c(0) \approx \SI{31,3}{mT}$ and $T_c\approx\SI{4,04}{K}$ for Sn, which are both in good agreement with literature \cite{onori_sn_crit_magn,kittel}. The decrease of $Q$ at fields of about \SI{80}{mT} is due to the remaining Pb resonator going normal conducting similar to above.

\section{Conclusion}
In this work we were able to access the electrodynamic properties of superconducting Pb microwave stripline resonators in variable magnetic field up to \SI{140}{mT} and in a temperature range between \SI{1,6}{K} and \SI{6,5}{K}. With the calculation of the penetration depth into the superconducting Pb and the surface resistance, the calculation of the complex conductivity was possible. This was done for the first time for a type~I superconductor with microwave stripline resonators in variable, parallel magnetic field and revealed similarities to temperature-dependent measurements, e.g. features similar in shape to a coherence peak.

Existence of remaining superconducting filaments at magnetic fields above the critical magnetic field of Pb has been found. The quality factor of the resonator, as well as the complex conductivity showed non-vanishing values at magnetic fields above the critical magnetic field and only vanished well above the critical magnetic field.
Combined with the gradual resistance increase only at $B>B_c(T)$ this indicates that our Pb specimen contained impurities making it a hard superconductor.

\textcolor{blue}{Determination} of the critical magnetic field and the critical temperature of Pb was possible, which have been found to be in good agreement with previous results. After exceeding the critical magnetic field once, hysteresis in the quality factor has been found on return to lower magnetic fields, as impurities in the resonator act as \textcolor{blue}{seeds} for normal conducting areas. These have been found to persist even in the superconducting phase, enhancing resistivity in the resonator, making it \textcolor{blue}{advisable} to always perform zero-field-cooling before using a Pb resonator in microwave spectroscopy studies in magnetic field. Alternatively, we showed the possibility to regain the original superconducting state by repeatedly sweeping the magnetic field to positive and negative value below $B_c$, enabling restoration \textcolor{blue}{of $Q$} even at very low temperatures.

By modifying the resonator to be able to study a Sn sample, we could demonstrate the possibility to examine superconducting samples with stripline resonators in a variable magnetic field. The critical magnetic field and the critical temperature of Sn could be determined and fit the BCS predictions well.

\section*{Acknowledgments}
We thank Gabriele Untereiner for the fabrication of the resonators.
Financial support by DFG and by Carl-Zeiss-Stiftung is thankfully acknowledged.\\
\newline
$\ddagger$ We estimate the alignment of the magnetic field to the resonator chip to be parallel within a maximum variance of \SI{1}{}-\SI{2}{\degree}.


\section*{References}
\begin{thebibliography}{unsrt}

\bibitem{dressel_buch} Dressel M and Gr{\"u}ner G 2002 \textit{Electrodynamics of Solids} (Cambridge University Press, Cambridge, UK)

\bibitem{Pracht2013} Pracht U S, Heintze E, Clauss C, Hafner D, Bek R, Werner W, Gelhorn S, Scheffler M, Dressel M, Sherman D, Gorshunov B, Il’in K S, Henrich D and Siegel M 2013
Electrodynamics of the Superconducting State in Ultra-Thin Films at THz Frequencies
\textit{IEEE Trans. THz Sci. Technol.} \textbf{3} 269

\bibitem{dressel_microwave_sigma} Donovan S, Klein O, Dressel M, Holczer K and Gr{\"u}ner G 1993 Microwave cavity perturbation technique: Part II: Experimental scheme \textit{Int J Infrared Millimeter Waves} \textbf{14} 12

\bibitem{kobayashi_microwave_q_sigma} Kobayashi Y, Imai T and Kayano H 1991 Microwave Measurement of Temperature and Current Dependences of Surface Impedance for High-Tc Superconductors \textit{IEEE Trans. Microw. Theory Techn.} \textbf{39} 9

\bibitem{Feller2002} Feller J R, Tsai C C, Ketterson J B, Smith J L and Sarma B K 2002
Evidence of electromagnetic absorption by collective modes in the heavy fermion superconductor UBe$_{13}$
\textit{Phys. Rev. Lett.} \textbf{88} 247005

\bibitem{Turner2003} Turner P J, Harris R, Kamal S, Hayden M E, Broun D M, Morgan D C, Hosseini A, Dosanjh P, Mullins G K, Preston J S, Liang R, Bonn D A and Hardy W N 2003 Observation of Weak-Limit Quasiparticle Scattering via Broadband Microwave Spectroscopy of a d-Wave Superconductor
\textit{Phys. Rev. Lett.} \textbf{90} 237005

\bibitem{Kitano2006} Kitano H, Ohashi T, Maeda A and  Tsukada I 2006 Critical microwave-conductivity fluctuations across the phase diagram of superconducting La$_{2 −- x}$Sr$_x$CuO$_4$ thin films
\textit{Phys. Rev. B} \textbf{73} 092504

\bibitem{Steinberg2008} Steinberg K, Scheffler M and Dressel M 2008 Quasiparticle response of superconducting aluminum to electromagnetic radiation
\textit{Phys. Rev. B} \textbf{77} 214517


\bibitem{Cea2014}Cea T, Bucheli D, Seibold G, Benfatto L, Lorenzana J, Castellani C 2014
Optical excitation of phase modes in strongly disordered superconductors
\textit{Phys. Rev. B} \textbf{89} 174506

\bibitem{Sherman2015} Sherman D, Pracht U S, Gorshunov B, Poran S, Jesudasan J, Chand M, Raychaudhuri P, Swanson M, Trivedi N, Auerbach A, Scheffler M, Frydman A, Dressel M 2015 The Higgs mode in disordered superconductors close to a quantum phase transition
\textit{Nature Phys.} \textbf{11} 188


\bibitem{Day2003} Day P K, LeDuc H G, Mazin B A, Vayonakis A and Zmuidzinas J 2003
A broadband superconducting detector suitable for use in large arrays
\textit{Nature} \textbf{425} 817

\textcolor{blue}{\bibitem{Wallraff2004} Wallraff A, Schuster D I, Blais A, Frunzio L, Huang R S, Majer J, Kumar S, Girvin S M and Schoelkopf R J 2004 Strong coupling of a single photon to a superconducting qubit using circuit quantum electrodynamics \textit{Nature} \textbf{431} 162-167}

\bibitem{Goeppl2008} G\"{o}ppl M, Fragner A, Baur M, Bianchetti R, Filipp S, Fink J M, Leek P J, Puebla G, Steffen L and Wallraff A 2008
Coplanar waveguide resonators for circuit quantum electrodynamics
\textit{J. Appl. Phys.} \textbf{104} 113904

\bibitem{Natarajan2012} Natarajan C M, Tanner M G and Hadfield R H 2012
Superconducting nanowire single-photon detectors: physics and applications
\textit{Supercond. Sci. Technol.} \textbf{25} 063001

\textcolor{blue}{\bibitem{OConnell2008} O'Connell A D, Ansmann M, Bialczak R C, Hofheinz M, Katz N, Lucero, McKenney C, Neeley M, Wang H, Weig E M, Cleland A N, Martinis J M 2008
Microwave Dielectric Loss at Single Photon Energies and milliKelvin Temperatures
\textit{Appl. Phys. Lett.} \textbf{92} 112903}

\textcolor{blue}{\bibitem{Barends2010} Barends R, Vercruyssen N, Endo A, de Visser P J, Zijlstra T, Klapwijk T M, Diener P, Yates S J C and Baselmans J J A 2010  Minimal resonator loss for circuit quantum electrodynamics \textit{Appl. Phys. Lett.} \textbf{97} 023508}
	
\textcolor{blue}{\bibitem{Macha2010} Macha P, van der Ploeg S H W, Oelsner G, Il'ichev E, Meyer H G, W{\"u}nsch S and Siegel M 2010 Losses in coplanar waveguide resonators at millikelvin temperatures\textit{Appl. Phys. Lett.} \textbf{96} 062503}
	
\textcolor{blue}{\bibitem{Megrant2012} Megrant A, Neill C, Barends R, Chiaro B, Yu Chen, Feigl L, Kelly J, Lucero E, Mariantoni M, O'Malley P J J, Sank D, Vainsencher A, Wenner J, White T C, Yin Y, Zhao J, Palmstrom C J, Martinis J M and Cleland A N 2012 Planar superconducting resonators with internal quality factors above one million \textit{Appl. Phys. Lett.} \textbf{100} 113510}

\textcolor{blue}{\bibitem{deVisser2013} de Visser P J, Baselmans J J A, Bueno J, Llombart N and Klapwijk T M 2014 Fluctuations in the electron system of a superconductor exposed to a photon flux
\textit{Nature Commun.} \textbf{5} 3130}



\bibitem{Schwartz2000} Schwartz A, Scheffler M and Anlage S M 2000
Determination of the magnetization scaling exponent for single-crystal La$_{0.8}$Sr$_{0.2}$MnO$_{3}$ by broadband microwave surface impedance measurements
\textit{Phys. Rev. B} \textbf{61} R870


\bibitem{Schwarze2015} Schwarze T, Waizner J, Garst M, Bauer A,	Stasinopoulos I, Berger H, Pfleiderer C and Grundler D 2015
Universal helimagnon and skyrmion excitations in metallic, semiconducting and insulating chiral magnets
\textit{Nat. Mater.} \textbf{14} 478




\bibitem{scheffler_heavy_fermions} Scheffler M, Schlegel K, Clauss C, Hafner D, Fella C, Dressel M, Jourdan M, Sichelschmidt J, Krellner C, Geibel C and Steglich F 2013 Microwave spectroscopy on heavy-fermion systems: Probing the dynamics of charges and magnetic moments \textit{Phys. Status Solidi B} \textbf{250} 439

\bibitem{Awasthi1993} Awasthi A M, Degiorgi L, Gr\"{u}ner G, Dalichaouch Y and Maple M B 1993
Complete optical spectrum of CeAl$_3$
\textit{Phys. Rev. B} \textbf{48} 10692

\bibitem{Scheffler2005b} Scheffler M, Dressel M, Jourdan M, Adrian H 2005
Extremely slow Drude relaxation of correlated electrons
\textit{Nature} \textbf{438} 1135

\bibitem{scheffler_heavy_fermions_2} Scheffler M, Dressel M and Jourdan M 2010 Microwave conductivity of heavy fermions in UPd$_{2}$Al$_{3}$ \textit{Eur. Phys. J. B} \textbf{74} 331


\bibitem{Hashimoto2009} Hashimoto K, Shibauchi T, Kato T, Ikada K, Okazaki R, Shishido H, Ishikado M, Kito H, Iyo A, Eisaki H, Shamoto S and Matsuda Y 2009
Microwave Penetration Depth and Quasiparticle Conductivity of PrFeAsO$_{1-y}$ Single Crystals: Evidence for a Full-Gap Superconductor
\textit{Phys. Rev. Lett.} \textbf{102} 017002

\bibitem{Truncik2013} Truncik C J S, Huttema W A, Turner P J, \"{O}zcan S, Murphy N C, Carri\`{e}re P R, Thewalt E, Morse K J, Koenig A J, Sarrao J L, Broun D M 2013
Nodal quasiparticle dynamics in the heavy fermion superconductor CeCoIn$_5$ revealed by precision microwave spectroscopy
\textit{Nat. Commun.} \textbf{4} 2477

\bibitem{Liu2013} Liu W, Pan L, Wen J, Kim M, Sambandamurthy G and Armitage N P 2013
Microwave Spectroscopy Evidence of Superconducting Pairing in the Magnetic-Field-Induced Metallic State of InOx Films at Zero Temperature
\textit{Phys. Rev. Lett.} \textbf{111} 067003

\textcolor{blue}{\bibitem{Beutel2016} Beutel M H, Ebensperger N G, Thiemann M, Untereiner G, Fritz V, Javaheri M, N{\"a}gele J, R{\"ö}sslhuber R, Dressel M and Scheffler M 2016 Microwave study of superconducting Sn films above and below percolation \textit{Supercond. Sci. Technol.} \textbf{29} 085011}


\bibitem{Scheffler2015} Scheffler M, Felger M M, Thiemann M, Hafner D, Schlegel K, Dressel M, Ilin K S, Siegel M, Seiro S, Geibel C and Steglich F 2015
Broadband Corbino spectroscopy and stripline resonators to study the microwave properties of superconductors \textit{Acta IMEKO} \textbf{4} 47


\bibitem{Booth1994}Booth J C, Wu D H and Anlage S M 1994
A broadband method for the measurement of the surface impedance of thin films at microwave frequencies
\textit{Rev. Sci. Instrum.} \textbf{65} 2082

\bibitem{Turner2004} Turner P J, Broun D M, Kamal S, Hayden M E, Bobowski J S , Harris R, Morgan D C, Preston J S, Bonn D A, Hardy W N 2004
Bolometric technique for high-resolution broadband microwave spectroscopy of ultra-low-loss samples
\textit{Rev. Sci. Instrum.} \textbf{75} 124

\bibitem{Scheffler2005a} Scheffler M and Dressel M 2005 Broadband microwave spectroscopy in Corbino geometry for temperatures down to 1.7 K
\textit{Rev. Sci. Instrum.} \textbf{76} 074702

\bibitem{Huttema2006} Huttema W A, Morgan B, Turner P J, Hardy W N, Zhou X, Bonn D A, Liang R, Broun D M 2006
Apparatus for high-resolution microwave spectroscopy in strong magnetic fields
\textit{Rev. Sci. Instrum.} \textbf{77} 023901


\bibitem{Pompeo2007} Pompeo N, Marcon R and Silva E 2007
Dielectric Resonators for the Measurement of Superconductor Thin Films Surface Impedance in Magnetic Fields at High Microwave Frequencies
\textit{J. Supercond.} \textbf{20} 71


\bibitem{DiIorio1988} DiIorio M S, Anderson A C, Tsaur B Y 1988
rf surface resistance of Y-Ba-Cu-O thin films
\textit{Phys. Rev. B.} \textbf{38} 7019

\bibitem{scheffler_stripline} Scheffler M, Fella C and Dressel M 2012 Stripline resonators for cryogenic microwave spectroscopy on metals and superconductors \textit{J. Phys.: Conf. Ser.} \textbf{400} 052031

\bibitem{hafner} Hafner D, Dressel M and Scheffler M 2014 Surface-resistance measurements using superconducting stripline resonators \textit{Rev. Sci. Instrum.} \textbf{85} 014702

\bibitem{kittel} Kittel C 2005 \textit{Introduction to Solid State Physics} (John Wiley \& Sons, Inc., Hoboken, NJ, USA)



\bibitem{Sichelschmidt2010} Sichelschmidt J, Kambe T, Fazlishanov I, Zakharov D, Krug von Nidda H-A, Wykhoff J, Skvortsova A, Belov S, Kutuzov A, Kochelaev B I, Pashchenko V, Lang M, Krellner C, Geibel C and Steglich F 2010
Low temperature properties of the electron spin resonance in YbRh$_2$Si$_2$
\textit{Phys. Status Solidi B} \textbf{250} 747


\bibitem{clauss_esr} Clauss C, Bothner D, Koelle D, Kleiner R, Bogani L, Scheffler M and Dressel M 2013 Broadband electron spin resonance from 500 MHz to 40 GHz using superconducting coplanar waveguides \textit{Appl. Phys. Lett.} \textbf{102} 162601

\bibitem{wiemann_esr} Wiemann Y, Simmendinger J, Clauss C, Bogani L, Bothner D, Koelle D, Kleiner R, Dressel M and Scheffler M 2015 Observing electron spin resonance between 0.1 and 67 GHz at temperatures between 50 mK and 300 K using broadband metallic coplanar waveguides \textit{Appl. Phys. Lett.} \textbf{106} 193505



\bibitem{Song2009} Song C, DeFeo M P, Yu K and Plourde B L T 2009
Reducing microwave loss in superconducting resonators due to trapped vortices
\textit{Appl. Phys. Lett.} \textbf{95} 232501

\bibitem{Bothner2012}Bothner D, Clauss C, Koroknay E, Kemmler M, Gaber T, Jetter M, Scheffler M, Michler P, Dressel M, Koelle D and Kleiner R 2012
Reducing vortex losses in superconducting microwave resonators with microsphere patterned antidot arrays
\textit{Appl. Phys. Lett.} \textbf{100} 012601

\bibitem{bothner_demagnitization} Bothner D, Gaber T, Kemmler M, Koelle D, Kleiner R, W{\"u}nsch S and Siegel M 2012 Magnetic hysteresis effects in superconducting coplanar microwave resonators \textit{Phys. Rev. B} \textbf{86} 014517

\bibitem{steele_molybden} Singh V, Schneider B H, Bosman S J, Merkx E P J Steele G A 2014 Molybdenum-rhenium alloy based high-Q superconducting microwave resonators \textit{Appl. Phys. Lett.} \textbf{105} 222601

\bibitem{song_vortices_in_re_al} Song C, Heitmann T, DeFeo M, Yu K, McDermott R, Neeley M, Martinis J and Plourde B 2009 Microwave response of vortices in superconducting thin films of Re and Al \textit{Phys. Rev. B} \textbf{79} 174512 

\bibitem{ge_flux_patterin_in_lead} Ge J, Gutierrez J, Raes B, Cuppens J and Moshchalkov V 2013 Flux pattern transitions in the intermediate state of a type-I superconductor driven by an ac field \textit{New J. Phys.} \textbf{15} 033013

\bibitem{alers_intermediate_state_pb} Alers P B 1957 Structure of the intermediate state in superconducting lead \textit{Phys. Rev.} \textbf{105} 104

\bibitem{velez_magnetic_lead} Velez S, Panades-Guinart C, Abril G, Garcia-Santiago A, Hernandez J and Tejada J 2008 Topological magnetic irreversibility in superconducting Pb samples of various shapes \textit{Phys. Rev. B} \textbf{78} 134501

\bibitem{velez_temp_magn_lead} Velez S, Garcia-Santiago A, Hernandez J and Tejada J 2012 The role of temperature in the magnetic irreversibility of type-I Pb superconductors \textit{J. Phys.: Condens. Matter} \textbf{24} 485701

\bibitem{schoenberg_superconductivity_book}{Schoenberg D 1938 \textit{Superconductivity} (Cambridge University Press, Cambridge, UK)}

\bibitem{seraphim_resistance_in_Ta}{Seraphim D P and Connell R A 1959 Resistance transitions in superconducting tantalum \textit{Phys. Rev.} \textbf{116} 3}

\bibitem{budnick_resistance_in_tantalum}{Budnick J I 1960 Some studies of the superconducting transition in purified tantalum \textit{Phys. Rev.} \textbf{119} 5}

\bibitem{cardona_surface_impedance} Cardona M, Fischer G and Rosenblum B 1963 Microwave surface impedance of superconductors of the second kind: In-Bi alloys \textit{Phys. Rev. Lett.} \textbf{12} 101

\textcolor{blue}{\bibitem{Schuster2005} Schuster D I, Wallraff A, Blais A, Frunzio L, Huang R-S, Majer J, Girvin S M and Schoelkopf R J 2005
ac Stark Shift and Dephasing of a Superconducting Qubit Strongly Coupled to a Cavity Field
\textit{Phys. Rev. Lett.} \textbf{94} 123602}

\textcolor{blue}{\bibitem{Chin1992} Chin C C, Oates D E, Dresselhaus G and Dresselhaus M S 1992
Nonlinear electrodynamics of superconducting NbN and Nb thin films at microwave frequencies
\textit{Phys. Rev. B} \textbf{45} 4788}

\bibitem{petersan_resonator_general} Petersan P and Anlage S 1998 Measurement of resonant frequency and quality factor of microwave resonators: Comparison of methods \textit{J. Appl. Phys.} \textbf{84} 6

\bibitem{wheeler} Wheeler H 1978 Transmission-Line Properties of a Strip Line Between Parallel Planes \textit{IEEE Trans. Microw. Theory Techn.} \textbf{26} 11

\bibitem{reeber_magnetic_field_orientation}{Reeber M D 1959 Geometric effects in the superconducting transition of thin films \textit{IBM J Res. Dev.} \textbf{3} 2}

\bibitem{koepke_neustes} K{\"o}pke M and Weis J 2014 Superconducting Pb as material for coplanar waveguide resonators on GaAs substrates \textit{Phys. C} \textbf{506} 143-145

\bibitem{saint-james_hc3_theoretical_prediction}{Saint-James D and Gennes P G 1963 Onset of superconductivity in decreasing fields \textit{Phys. Lett.} \textbf{7} 5}

\bibitem{cardona_surface_supercond} Cardona M and Rosenblum B 1964 Microwave observation of superconductivity above the upper critical field \textit{Phys. Lett.} \textbf{8} 5

\bibitem{hempstead_hc3_in_typeII}{Hempstead C F and Kim Y B 1964 Resistive transitions and surface effects in type-II superconductors \textit{Phys. Rev. Lett.} \textbf{12} 145}

\bibitem{crescenzo_pb_magn_temp_thickness} Di Crescenzo E, Indovina P, Onori S and Rogani A 1973 Temperature and Thickness Dependence of Critical Magnetic Fields in Lead Superconducting Films \textit{Phys. Rev. B} \textbf{7} 7

\bibitem{cody_pb_magn_thickness} Cody G D and Miller R E 1968 Magnetic transitions of superconducting thin films and foils. I. Lead \textit{Phys. Rev.} \textbf{173} 481

\bibitem{indovina_pb_crit_magn_microwave} Indovina P, Onori S and Tabet E 1970 Magnetic Field Modulation of the Microwave Impedance of Pb Superconducting Films \textit{Solid State Commun.} \textbf{8} 21

\bibitem{desorbo_intermediate_state}{DeSorbo W and Healy W A 1964 The intermediate state of some superconductors? \textit{Cryogenics} \textbf{4} 5}

\bibitem{bcs_theory} Bardeen J, Cooper N and Schrieffer J 1957 Theory of Superconductivity \textit{Phys. Rev.} \textbf{108} 5

\bibitem{tinkham_introduction} Tinkham M 2004 \textit{Introduction to Superconductivity: Second Edition} (Dover Publications, Mineola, NY, USA)

\bibitem{glaever_energy_gap} Giaever I 1960 Energy gap in superconductors measured by electron tunneling \textit{Phys. Rev. Lett.} \textbf{5} 4

\bibitem{richards_energy_gap} Richards P L and Tinkham M 1960 Far-infrared energy gap measurements in bulk superconducting In, Sn, Hg, Ta, V, Pb, and Nb. \textit{Phys. Rev.} \textbf{119} 2

\bibitem{dressel_edyn_supercond} Dressel M 2013 Electrodynamics of Metallic Superconductors \textit{Adv. Cond. Matter Phys.} \textbf{2013} 104379

\bibitem{holczer_coherence_peak} Holczer K, Klein O and Gr{\"u}ner G 1991 Observation of the conductivity coherence peak in superconducting Pb \textit{Solid State Commun.} \textbf{78} 10

\bibitem{thiemann_sigma} Thiemann M, Bothner D, Koelle D, Kleiner R, Dressel M and Scheffler M 2014 Niobium stripline resonators for microwave studies on superconductors \textit{J Phys Conf Ser.} \textbf{568} 022043

\bibitem{onori_sn_crit_magn} Onori S and Rogani A 1980 Critical magnetic fields in tin superconducting films \textit{Physica B+C} \textbf{100} 1

\end{thebibliography}
\end{document}